# Dimensional cross-over of the bandgap transition in quasi-two-dimensional MoS$_2$

Jinhua Hong[1‡], Kun Li[2‡], Chuanhong Jin[1*], Xixiang Zhang[3*], Ze Zhang[1], Jun Yuan[1,4*]

[1]State Key Laboratory of Silicon Materials, Key Laboratory of Advanced Materials and Applications for Batteries of Zhejiang Province, Department of Materials Science and Engineering, Zhejiang University, Hangzhou, Zhejiang 310027, P. R. China

[2]Advanced Nanofabrication, Imaging and Characterisation Core Lab, King Abdullah University of Science and Technology (KAUST), Thuwal 239955, Kingdom of Saudi Arabia

[3]Division of Physical Science, King Abdullah University of Science and Technology (KAUST), Thuwal 239955, Kingdom of Saudi Arabia

[4]Department of Physics, University of York, Heslington, York, YO10 5DD, UK

**Abstract:**

**The anisotropy of the electronic transition is an important physical property not only determining the materials' optical property, but also revealing the underlying character of the electronic states involved. Here we used momentum-resolved electron energy-loss spectroscopy to study the evolution of the anisotropy of the electronic transition involving the low energy valence electrons in the free-standing MoS$_2$ systems as the layer thickness was reduced to monolayer. We used the orientation and the spectral-density analysis to show that indirect to direct band-gap transition is accompanied by a three- to two-dimensional anisotropy cross-over. The result provides a logical explanation for the large sensitivity of indirect transition to the change of thickness compared with that for direct transition. By tracking the energy of indirect transition, we also revealed the asymmetric response of the valence band and conduction band to the quantum confinement effect. Our results have implication for future optoelectronic applications of atomic thin MoS$_2$.**

Atomically thin molybdenum disulfide (MoS$_2$), as a representative member of the emerging 2D transition metal dichalcogenides (TMDs) [1,2], has attracted intensive research efforts owing to its unique structure, optoelectronics [3-8] and valleytronics [9,10]. Electronic structure, including its anisotropy, holds the key to the understanding of those interesting properties. Most important for atomically thin $MoS_2$ is the degree of anisotropy of the electronic structures and their evolution as the layer number increases. However, there has been very little experimental work as it is particularly difficult to measure optical transition with an out-of-plane polarisation for atomically thin $MoS_2$. Through probing the anisotropic nature of the transition, we can study the underlying character of the electronic states involved and understand how best to excite these transitions. Also receiving a lot of attention is the prediction of the transition from indirect to direct bandgap when the thickness of $MoS_2$ is reduced to monolayer by *ab initio* theory [11,12] and indirect confirmation by the observation of strong photoluminescence enhancement in the monolayer [13,14]. The strong chiral pumping effect observed in the bandgap absorption [9,10] has opened up the possibility of dynamical control of valley specific carrier density in the monolayer system which lacks inversion symmetry. Given that, it is important to obtain direct experimental evidence for the nature of bandgap transition and their evolution with the number of layers. Very recently, the valence band has been probed in 2D TMDs to elucidate its evolution with the thickness using angle-resolved photoemission spectroscopy (ARPES) [15,16]. However, ARPES is not sensitive to the unoccupied states in undoped semiconductors and may also be influenced by the presence of the substrate [15].

In this work, we have carried out the first momentum-dependent electron energy-loss spectroscopy (EELS) characterisation on the electronic structures of atomically thin $MoS_2$ (by mechanical exfoliation) to reveal the anisotropic properties of the electronic excitations. As an alternative tool to probe electronic structures of semiconductors, EELS has been used to study the valence electron excitation such as bandgap transition [17,18] and plasmonics [19,20]. Compared to optical methods, EELS has its specific advantage in terms of the energy and momentum range accessible and in its unrivalled high spatial resolution. The adjustable momentum transfer in EELS is particularly suited to probe the anisotropy of the electronic transitions because the direction of the momentum transfer ranges from parallel to perpendicular to electron beam at the characteristic scattering angle ($\theta_E = E/2E_0$) [21]. In addition, we have extended a power-law analysis of the spectral intensity near the bandgap to two dimensional systems and show not only that the direct evidence for the transition from indirect to direct bandgap excitation as the thickness is reduced to monolayer, but also that this is accompanied by a cross-over from three-dimensional indirect transition to a two-dimensional direct transition.

The monolayer MoS$_2$ layer has a graphite-like structure with contrasting sublattices as shown in Fig. 1(a). Fig. 1(b) sketches its reciprocal space with the first Brillouin zone (FBZ) marked in dashed line. Fig. 1(c) shows the diffraction pattern of a free-standing monolayer MoS$_2$ along the c-axis. Superimposed on the diffraction pattern is a green ring marking the size of the round spectrometer entrance aperture (SEA) used in the measurement. A rotation holder was used to orient the sample to align the specific in-plane orientation (for instance ΓK direction in Fig. 1(c)) along the energy dispersing direction (defined as the q$_x$ direction in the momentum space). Free-standing areas of high quality MoS$_2$ monolayer were selected to avoid the substrate effect. The q-E diagram shown in Fig. 1(d) is obtained by summation of 200 individual 1.0 second drift-corrected measurements to enhance the signal-to-noise ratio. The momentum-dependent spectra are line plots as a function of q$_y$ as shown in Fig. 1(d). Each line plot is an intensity integration along the q$_x$ direction (the energy dispersing direction) over the momentum transfer range limited by SEA. For a uniaxial crystal, the experimental intensity can be written as [22]:

$$\int_{-\sqrt{q_0^2-q_y^2}}^{\sqrt{q_0^2-q_y^2}} \frac{\text{Im}(\varepsilon_{//})q_E^2 + \text{Im}(\varepsilon_\perp)(q_x^2+q_y^2)}{\left|\varepsilon_{//}q_E^2 + \varepsilon_\perp(q_x^2+q_y^2)\right|^2} dq_x \Delta q_y \quad (1)$$

where $q_0$ (=0.54Å$^{-1}$) is the size of the SEA. Even in this partially momentum-integrated form, the anisotropy information should still be visible as we demonstrate below.

Figure 2(a) shows the momentum-dependent EEL spectra with q$_y$ along the ΓK direction in monolayer. Only excitations up to 5 eV are displayed because of our focus on the electronic structure near the bandgap region. Plasmon excitation at higher energy losses will be discussed in a separate paper. Three noticeable features are observed: a weak transition at 2.0 eV (marked by A,B), a group of strong transition centered at 3.1 eV, and a broad peak at 4.5 eV. A similar result for EELS along the ΓM direction is also obtained [23], indicating that it is a good approximation to treat the dielectric response to be uniaxial as described by Eqn.(1).

The momentum-dependent EEL spectra show strong evidence for the anisotropy of the electronic transition in the energy range studied. In Fig. 2(b), we have plotted the intensity variations of the three dominant absorption bands with the momentum transfer $q_y$, normalised to their values at $q_y=0$. We expect that the contribution from the out-of-plane component ($q // c$) to dominate the spectrum at $q_y=0$ and that of the in-plane ($q \perp c$) to dominate at large $q_y$, although the integration over $q_x$ complicates the spectra interpretation a little bit, as it mixes the in-plane contribution into the $q_y=0$ spectrum. As a result, two different behaviors for the in-plane and the out-of-plane components are observed in Fig. 2(b) as expected. For the A,B peak at 2eV, the normalised intensity is initially independent of $q_y$, in agreement with the expected dependence of the in-plane contribution [24]. On the other hand, the intensities of the broad α, β peaks centered at 3.1 eV and 4.5 eV drop more rapidly, consistent with the contribution from the out-of-plane excitation which occurs at the small momentum transfer [24].

Another sign for the strong orientation effect of the dielectric response is the systematic disappearance or appearance of certain spectral features, such as the α peak or the δ peak (at 3.9 eV), as $q_y$ increases [Fig. 2(a)]. We will make simple assumption that for the small energy range considered, the integrated spectra are linear superposition of the in-plane and out-of-plane components. This is based on the observation that the sharp δ peak absent in the small angle is the sole property of the in-plane spectra and that the α peak is mostly present in the small scattering angle. The exact determination of in-plane and out-of-plane components is described in the supplemental material and the approximate results are shown in Fig 2(c). Here the out-of-plane ($q // c$) components are similar to the electronic transition probed by light with $E // c$ polarisation, which is quite difficult to realise on atomic layers by optical means because of the transverse nature of the electromagnetic field.

The orientation-resolved spectra in monolayer MoS$_2$ immediately suggest that the bandgap in the out-of-plane direction has a rather higher value of 2.5 eV. This means that the direct-gap transition in monolayer seen around 2 eV regions has a pure 2D character. On the other hand, the in-plane and the out-of-plane transitions in the multilayer system near the bandgap region which occurs at the smaller energy of 1.5 eV seem to have much less orientation dependence, suggesting the electronic transition involved have a 3D character. Our result for the multilayer MoS$_2$ is consistent with the optical reflectivity from the bulk MoS$_2$ [25,26], and our result for the monolayer system is in agreement with the *ab initio* calculation by Kumar et al [27].

The bandgap transition in the monolayer MoS$_2$ presents a dispersionless character as shown in Fig. 2(d), illustrating the excitonic nature of this direct transition which has been also identified in the optical spectra [13]. Theoretically the valence band spin-splitting leads to a 0.15 eV difference between the spin-orbital split states of the excitons [28]. However, the phonon scattering and the limited energy resolution (0.14 eV) does not allow us to discriminate these two close direct-transition peaks. In fact, the other two sharp transition features we have identified all show negligible variation over the momentum range being probed, so we can ignore the dispersion effect in our discussions of momentum-dependent spectrum.

A further evidence of the three-dimensional character of the bandgap transition in the multilayers comes from fitting to the power-law (E-E$_g$)$^n$. Figure 3a shows the fine structures of low-loss spectra from MoS$_2$ with variable thickness. Here ultrasmall atomic thickness excludes the possible influence from Cerenkov loss. Brown and Rafferty [29] have shown that the transition matrix element can be considered to be constant over the energy range and the spectral intensity for a direct bandgap transition is determined by the joint density of states JDOS(E) alone and that for indirect bandgap is given by (E-E$_g$)×JDOS(E). For the 3D semiconductors they have considered, the JDOS for a parabolic dispersion follows (E-E$_g$)$^{0.5}$. Hence in the 3D case, direct bandgap transition has an exponent of n=0.5 and indirect transition n=1.5. They have verified their results experimentally for a number of 3D semiconducting materials.

We have extended their argument on the energy-dependence of the spectral intensity for direct and indirect bandgap transitions to two dimensional systems and our analysis suggests that the corresponding exponent would be n=0 and n=1 respectively. The fitted exponents for indirect bandgap transitions in multilayers [Fig. 3(b)] are all close to 1.5, together with the orientation-insensitive bandgap transition in Fig. 2(c), demonstrating they have 3D character.

The electronic structure of MoS$_2$ is typical of transition metal dichalcogenides system, with the 4d orbitals situated within the larger energy σ-σ* gap of bonding and antibonding s-p orbitals [30,31]. Because of the trigonal prismatic nature of the S ligand atom arrangement, the 4d orbitals are further split into bonding e$_g$-like upper band involving $d_{xz}$, $d_{yz}$ orbitals and a lower t$_{2g}$-like lower band involving $d_{z^2}$, $d_{xy}$ and $d_{x^2-y^2}$ orbitals. The hybridization among the symmetry-allowed combination of d-orbitals produces the resulting electronic states, but at high symmetry points, it is useful to discuss the states and the electronic transitions in terms of the atomic orbitals involved. *Ab initio* calculations have identified the direct bandgap as originating from the transition at the K-point of the FBZ which has a predominant Mo $d_{x^2-y^2}$ character, and the indirect bandgap transition arise from transition from the local valence band maximum at the Γ-point to local conduction band valley Q-point [in Fig. 3(c)]. The electronic orbital character of the valence band extreme at the Γ-point has been identified with Mo-$d_{z^2}$-S-$p_z$ hybrid [30]. The dipole-allowed transition detected by EELS requires a parity change, thus the direct d-d transition at the K-point should strictly only be allowed for $E \perp c$ polarisation as we have observed for the monolayer case and that of the indirect transition from the Γ-point should only be allowed in $E // c$ polarisation. Our result suggests that these selection rules are relaxed under the interlayer coupling. The three-dimensional nature of the indirect bandgap provides a logical explanation for the observed strong layer dependence of the indirect bandgap values [13]. The non-bonding nature of the $d_{x^2-y^2}$ orbital also means that the monolayer MoS$_2$ could be a more durable ultrathin photodetector [7,8] as it is less likely to suffer from photo-bleaching effect.

Using our power-law fitting, we can follow the evolution of the indirect bandgap as a function of the layer thickness, even after the cross-over from indirect-to-direct bandgap transition [Fig. 3(b)]. This is because the overlapping direct bandgap transition in the monolayer system should have a power-law dependence of (E-E$_g$)$^0$ similar to the JDOS of the two-dimensional character of the band structure near the K-point. Although near the threshold, MoS$_2$ does not follow such power-law due to the strong excitonic effect, we expect that such a power-law dependence prevails above the bandgap region where the excitonic effect is absent. Indeed, such a flat absorption band has been well known in the related TMDs [26]. We therefore identify the rising absorption above direct transition at 2 eV in monolayer with the same transition responsible for the indirect Γ→Q transition in multilayer MoS$_2$. Giving support to this identification, the power-fitting returns a spectral intensity exponent of

1.56±0.2. The deduced transition energy is 2.26±0.06 eV, corresponding to indirect transition in monolayer.

Figure 3d summarizes the transition threshold energies in $MoS_2$ with different thickness. Here, regarding the direct transition energies, they are denoted by the position of A,B peaks for different layers without distinguishing. Calculations by Kumar and Molina-Sanchez et al [27,32] predict downward shift of the VBM at Γ-point when the thickness is reduced. This has been collaborated by ARPES measurements although the bandwidth of the valence band is smaller in experiment because of the substrate effect [15]. To date, there is no experimental information about the movement of the conduction band structure, nor are DFT calculations reliable due to the well-known bandgap under-estimate problem. Therefore, we use our determination of the energies for the direct and indirect transitions [in Fig. 3(d)] to map out the movement of the conduction band valley at Q-point. The ARPES data suggests that the valence band maximum at the Γ-point was downward shifted from its bulk value to the monolayer by 0.7 eV. Further in ours, the evolution of the indirect transition energy suggests that the local valley located at the Q-point moves upward by a much smaller amount (0.3 eV). This may be related to different atomic orbital mixing at the valence band extreme Γ-point and that at the conduction band valley Q-point, as the latter has substantial hybridization between $d_{z^2}$ and $d_{xy}$, $d_{x^2-y^2}$ orbitals. This asymmetry will result in many physical properties, such as changes in the effective mass of electrons and holes which are important for the transport in few-layer $MoS_2$ systems.

Other sharp absorption peaks above bandgap absorption threshold can also be mapped to Von Hover's critical points in the joint density of sates, such as the 3.9 eV transition identified by other calculations [27]. Several calculated electronic structures of 1H-$MoS_2$ and a few layer 2H-$MoS_2$ indicate the α peak at 3.1 eV [Fig. 2(a)] should arise from the transition between the parallel conduction band and valence band around Q point, where high density of states is involved [32]. These critical points should be helpful in allowing us to check against the accuracy of the various calculated band structures which are the basis for further electronic, transport and optical study of these atomically thin materials.

In conclusion, we directly confirm a transition from the indirect gap to the direct gap as the thickness is reduced to monolayer, and also observe that indirect gap transition has a prominent 3D character and the direct bandgap in the monolayer system has a pure 2D character. The 3D character of the indirect transition explains its sensitivity to the thickness. In addition, in conjunction with the ARPES and band structure calculation, we are able to show that the local valence band maximum at the Γ-point is more sensitive to the interlayer coupling than the conduction band minimum at the Q-point in the FBZ. Our results allow us to quantitatively study the band structure, which are important to the optical, transport and possible valleytronic applications of atomically thin $MoS_2$.

**Acknowledgements**

This work is financially supported by the National Science Foundation of China (51222202), the National Basic Research Program of China (2014CB932500), the Program for Innovative Research Team in University of Ministry of Education of China (IRT13037) and the Fundamental Research Funds for the Central Universities (2014XZZX003-07). J.Y. acknowledges Pao Yu-Kong International Foundation for a Chair Professorship. Dr. He Tian is kindly acknowledged for a critical reading.

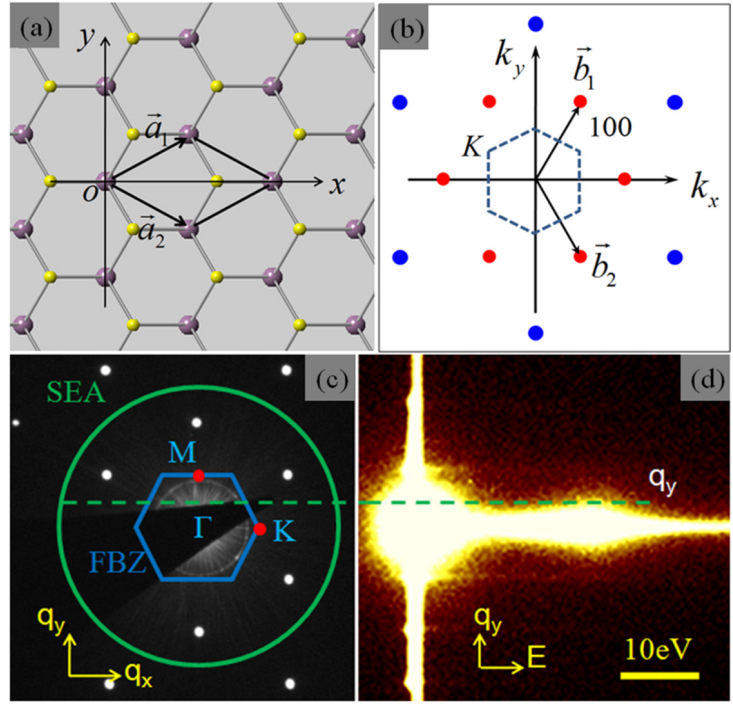

FIG. 1 (color online). (a) The c-axis projection of the primitive cell of monolayer MoS$_2$ is displayed with two lattice basis vector **a**$_1$ and **a**$_2$. The atoms are color coded (purple for Mo and yellow for S$_2$). (b) The reciprocal lattice with basic vector **b**$_1$, **b**$_2$ together with the FBZ outlined in dashed lines. (c) The scattering geometry of momentum resolved EELS in momentum space limited by a 2.5 mm diameter spectrometer entrance aperture (SEA). The SEA covers only (100) diffraction spot and its symmetric counterparts. (d) Corresponding q-E diagram with q$_y$ along the ΓM direction in (c).

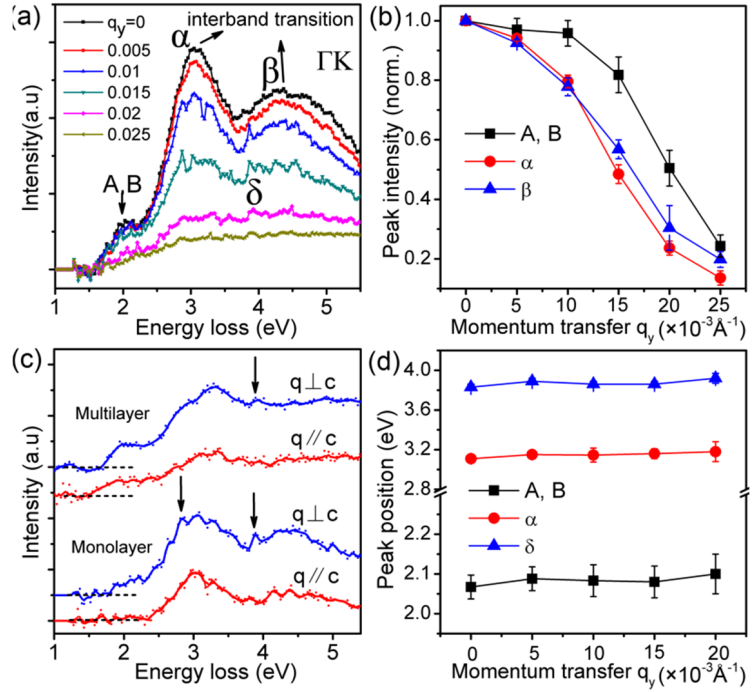

FIG. 2 (color online). (a) Momentum-resolved low loss spectra with q$_y$ in ΓK direction. The unit of q is Å$^{-1}$. The weak peak at 2.0 eV is due to direct transition of A, B excitons, and the strong peaks at 3.1 eV and 4.5 eV result from strong inter-band transitions. (b) The relative change of A,B, α, β peak intensities around 2 eV, 3.1 eV and 4.5 eV with momentum transfer $q_y$. Peak intensity at $q_y$ is normalised by its counterpart at $q_y=0$. (c) In-plane and out-of-plane absorption of monolayer and multilayer $MoS_2$. (d) The peak energy (peak position) of major absorption peaks in (a).

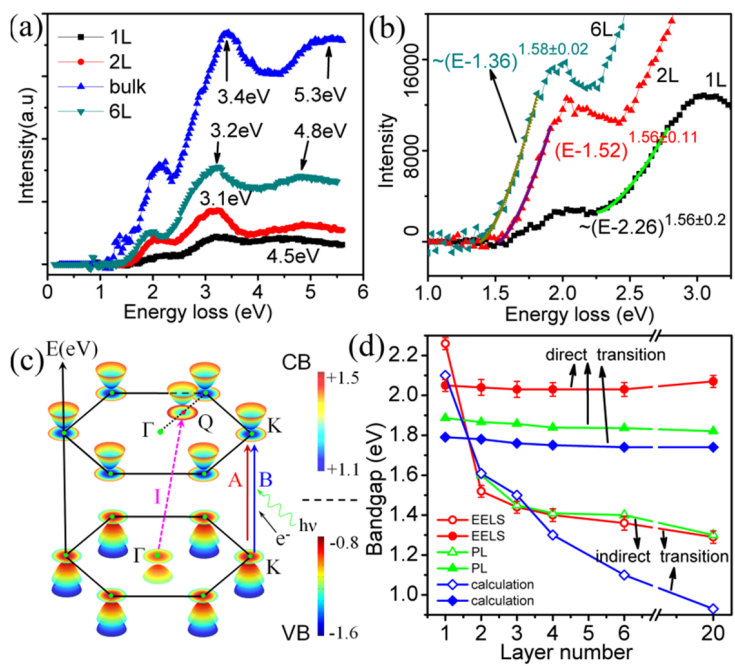

FIG. 3 (color online). Thickness-dependent bandgap evolution with a dimensional crossover. (a) The inter-band transition in $MoS_2$ with different layers. (b) A closer look at band edge transition. Nonlinear fittings of near gap fine structures give the indirect transition energy (Γ→Q transition). (c) Schematic illustration of band structure of monolayer $MoS_2$ with global conduction band minimum (CBM) and valence band maximum (VBM) both at K point. The second VBM is located at Γ-point and the second CBM at Q-point (almost midpoint of straight line ΓK). For the sake of brevity, only one Q point paraboloid is drawn (altogether six). The false color discs are the projection of parabolic dispersion to describe the positions of local VBM or CBM. (d) A summarized bandgap data between this work, and other reports [13,14,28,33]. Note the direct gaps of different layers are extracted from peak energy of A, B exciton transitions.

‡ These authors contributed equally to this work.

Corresponding authors.

chhjin@zju.edu.cn, xixiang.zhang@kaust.edu.sa, jun.yuan@york.ac.uk.

## References

[1] Q. H. Wang, K. Kalantar-Zadeh, A. Kis, J. N. Coleman, and M. S. Strano, Nature Nanotechnology **7**, 699 (2012).
[2] C. Ataca, H. Sahin, and S. Ciraci, Journal of Physical Chemistry C **116**, 8983 (2012).
[3] Y. Zhang, J. Ye, Y. Matsuhashi, and Y. Iwasa, Nano Letters **12**, 1136 (2012).
[4] B. Radisavljevic, A. Radenovic, J. Brivio, V. Giacometti, and A. Kis, Nature Nanotechnology **6**, 147 (2011).
[5] B. Radisavljevic, M. B. Whitwick, and A. Kis, ACS Nano **5**, 9934 (2011).
[6] B. W. H. Baugher, H. O. H. Churchill, Y. Yang, and P. Jarillo-Herrero, Nano Letters **13**, 4212 (2013).
[7] Z. Yin *et al.*, ACS Nano **6**, 74 (2012).
[8] H. S. Lee, S.-W. Min, Y.-G. Chang, M. K. Park, T. Nam, H. Kim, J. H. Kim, S. Ryu, and S. Im, Nano Letters **12**, 3695 (2012).
[9] H. Zeng, J. Dai, W. Yao, D. Xiao, and X. Cui, Nature Nanotechnology **7**, 490 (2012).
[10] K. F. Mak, K. He, J. Shan, and T. F. Heinz, Nature Nanotechnology **7**, 494 (2012).
[11] T. Li and G. Galli, Journal of Physical Chemistry C **111**, 16192 (2007).
[12] S. Lebegue and O. Eriksson, Physical Review B **79**, 115409 (2009).
[13] K. F. Mak, C. Lee, J. Hone, J. Shan, and T. F. Heinz, Physical Review Letters **105**, 136805 (2010).
[14] A. Splendiani, L. Sun, Y. Zhang, T. Li, J. Kim, C.-Y. Chim, G. Galli, and F. Wang, Nano Letters **10**, 1271 (2010).
[15] W. Jin *et al.*, Physical Review Letters **111**, 106801 (2013).
[16] Y. Zhang *et al.*, Nature Nanotechnology **9**, 111 (2014).
[17] L. Gu *et al.*, Physical Review B **75**, 195214 (2007).
[18] R. Arenal, O. Stephan, M. Kociak, D. Taverna, A. Loiseau, and C. Colliex, Physical Review Letters **95**, 127601 (2005).
[19] W. Zhou, J. Lee, J. Nanda, S. T. Pantelides, S. J. Pennycook, and J.-C. Idrobo, Nature Nanotechnology **7**, 161 (2012).
[20] P. Wachsmuth, R. Hambach, M. K. Kinyanjui, M. Guzzo, G. Benner, and U. Kaiser, Physical Review B **88**, 075433 (2013).
[21] R. F. Egerton, *Electron Energy-Loss Spectroscopy in the Electron Microscope* (Plenum Press, New York, 1996), 3 edn.
[22] Y. Sun and J. Yuan, Physical Review B **71**, 125109 (2005).
[23] See Supplemental Material Fig. S6.
[24] Detailed mathematical derivation is provided in Supplemental Material and see

Fig. S9.

[25] A. R. Beal and H. P. Hughes, Journal of Physics C-Solid State Physics **12**, 881 (1979).
[26] W. Liang, Journal of Physics C: Solid State Physics **6** (1973).
[27] A. Kumar and P. K. Ahluwalia, Materials Chemistry and Physics **135**, 755 (2012).
[28] E. S. Kadantsev and P. Hawrylak, Solid State Communications **152**, 909 (2012).
[29] B. Rafferty and L. M. Brown, Physical Review B **58**, 10326 (1998).
[30] Mattheis.Lf, Physical Review B **8**, 3719 (1973).
[31] M. Chhowalla, H. S. Shin, G. Eda, L.-J. Li, K. P. Loh, and H. Zhang, Nature Chemistry **5**, 263 (2013).
[32] A. Molina-Sanchez, D. Sangalli, K. Hummer, A. Marini, and L. Wirtz, Physical Review B **88**, 045412 (2013).
[33] A. Kumar and P. K. Ahluwalia, European Physical Journal B **85**, 186 (2012).